\documentclass[conference]{IEEEtran}
\ifCLASSINFOpdf
  \usepackage[pdftex]{graphicx}
  \graphicspath{{../pdf/}{../jpeg/}{./figure/}}
  \DeclareGraphicsExtensions{.pdf,.jpeg,.png}
\else
\fi
\usepackage[caption=false,font=footnotesize]{subfig}
\begin{document}
%
%
\title{Design and status of the Mu2e crystal calorimeter}



\author{\IEEEauthorblockN{ N. Atanov, V. Baranov, J. Budagov,  Yu. I. Davydov, V. Glagolev, V. Tereshchenko, 
Z. Usubov $                   $ }
\IEEEauthorblockA{Joint Institute for Nuclear Research, Dubna, Russia} \\
\and
\IEEEauthorblockN{ $   $ F. Cervelli,  S. Di Falco,  S. Donati, L. Morescalchi,  E. Pedreschi, G. Pezzullo,  F. Raffaelli,   F. Spinella}
\IEEEauthorblockA{INFN sezione di Pisa, Pisa, Italy }\\
\and
\IEEEauthorblockN{ $          $ F. Colao, M. Cordelli, G. Corradi, E. Diociaiuti,  R. Donghia, S. Giovannella, F. Happacher, M. Martini, \\ S. Miscetti$^{*}$, M. Ricci, A. Saputi, I. Sarra}\IEEEauthorblockA{ Laboratori Nazionali di Frascati dell' INFN, Frascati, Italy\\ ($*$) email: Stefano.Miscetti@LNF.INFN.IT}\\
\IEEEauthorblockN{ B. Echenard, D. G. Hitlin, T. Miyashita, F. Porter, R. Y. Zhu} 
\IEEEauthorblockA{ California Institute of Technology, Pasadena, USA} \\
\IEEEauthorblockN{ F. Grancagnolo, G. Tassielli}
\IEEEauthorblockA{ INFN sezione di Lecce, Lecce, Italy }\\
\IEEEauthorblockN{ P. Murat}
\IEEEauthorblockA{ Fermi National Accelerator Laboratory, Batavia, Illinois, USA }\\}


%

\maketitle

\begin{abstract}
The Mu2e experiment at Fermilab searches for the charged-lepton flavour violating (CLFV) conversion of 
a negative muon into an electron in the field of an aluminum nucleus, with a distinctive signature of a mono-energetic electron of
energy slightly below the muon rest mass (104.967 MeV). The Mu2e goal is to improve by four orders of magnitude the search sensitivity
with respect to the previous experiments.  Any observation of a CLFV signal will be a clear indication of
new physics.

The Mu2e detector is composed of a tracker, an electromagnetic calorimeter and an external veto for cosmic rays surrounding the
solenoid. The calorimeter plays an important role in providing  particle identification capabilities, a fast online trigger filter,
a seed for track reconstruction while working in vacuum, in the presence of 1 T axial magnetic field and in an harsh radiation environment.  
The calorimeter requirements are to provide a large acceptance for 100 MeV electrons and reach at these energies:
(a) a time resolution better than 0.5 ns; 
(b)  an energy resolution $< 10\%$  and
(c)  a position resolution of 1 cm.

The calorimeter design consists of two disks, each one made of 674 undoped CsI crystals read  by two large area arrays of UV-extended
SiPMs.  We report here the construction and test of the Module-0 prototype.
The Module-0 has been exposed to an electron
beam in the energy range around 100 MeV at the Beam Test Facility in Frascati. Preliminary results of timing and energy
resolution at normal incidence are shown. A discussion of the technical aspects of the calorimeter engineering is also reported
in this paper.

\end{abstract}


%
\IEEEpeerreviewmaketitle

\section{Introduction}
The Mu2e experiment~\cite{mu2e} at Fermilab searches for the charged-lepton flavour violating (CLFV) neutrino-less conversion of a negative muon 
into an electron in the field of an aluminum nucleus. The dynamics of such a  process is well modelled by a two-body decay, resulting 
in a mono-energetic conversion electron (CE) of energy slightly below the muon rest mass (104.967 MeV). CLFV processes  in
the muon channels are forbidden in the Standard Model (SM) and remain completely negligible, BR($\mu \to$ e $\gamma) = 10^{-52}$~\cite{bilenky},  
even assuming neutrino oscillations. Observation of CLFV candidates is a clear indication of new physics beyond
the Standard Model~\cite{newphysics}. If there is no $\mu\to$ e conversion, the average 90 \%  upper limit on the ratio between the conversion 
and the capture rates ($R_{\mu e}$) is $ < 8\ \times\ 10^{-17}$, thus improving the current
limit~\cite{sindrum-ii} by four orders of magnitude.
 
\begin{figure}[!t]
\centering
\includegraphics[width=3.6 in]{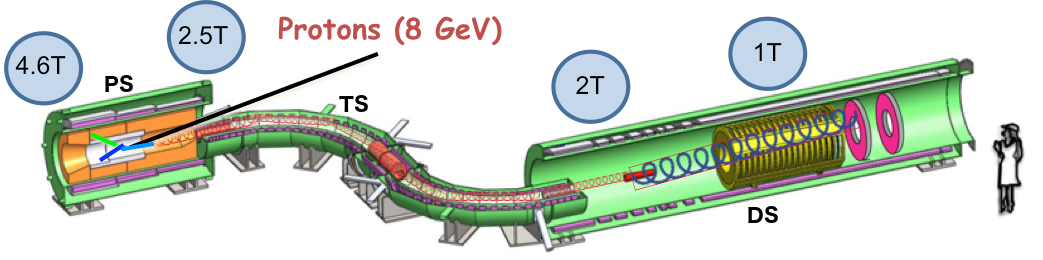}
\caption{Layout of the Mu2e experiment. The PS, DS and TS solenoids are indicated in the picture. The Cosmic Ray Veto, surrounding the DS and part of the TS solenoids, is not displayed.}
\label{mu2e_layout}
\end{figure}
In Fig. \ref{mu2e_layout}, the layout of the Mu2e experiment is shown. A large solenoidal
system is used to produce and transport a negative muon beam to an aluminum target.
A pulsed beam of 8 GeV protons is sent on a tungsten target inside the Production Solenoid (PS)
to produce low momentum pions and muons that are then funnelled by the graded field inside the
S-shaped Transport Solenoid (TS). Here, the pions decay to muons and a charge selection
is performed by means of a middle section collimation system. Finally, a very intense
pulsed negative muon beam ($ \sim10^{10} \mu/$sec) enters the Detector Solenoid (DS)
and is stopped on an aluminum target. In three years we expect to collect 6 $\times10^{17}$ muon
stops, sufficient  to reach our goal.
Decay products are analized by the tracker \cite{tracker} and
calorimeter \cite{calorimeter} systems. Cosmic ray muons can produce fake CE
candidates when interacting with the detector materials. In order to reduce their contributions
the external area of the DS, and a part of the TS, are covered by a Cosmic Ray Veto (CRV) \cite{CRV}
system.
\begin{figure}[!t]
\centering
\includegraphics[width=3.6in]{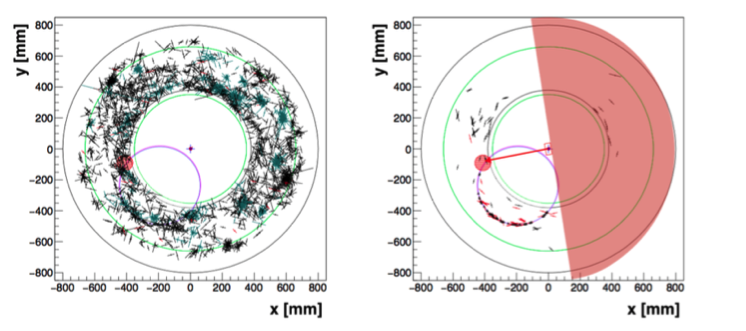}
\caption{Transversal view of the Mu2e detector and distribution of hits
on the tracker for one Monte Carlo signal event: (left) without any cuts and (right) with a cut on the timing between hits and the CE  calorimeter cluster.}
\label{caltrack}
\end{figure}

Around 50\% of the muon beam is stopped by the target while the rest
ends on the beam dump  at the end of the cryostat.
Muons stopped in the aluminum target  are captured in an atomic
excited state and promptly cascade to the 1S ground state, with 39\% decaying in
orbit (DIO) and the remaining 61\% captured by the nucleus. Low energy photons, neutrons and
protons  are emitted in the nuclear capture process. They constitute an
irreducible source of accidental activity that is the origin of  a large neutron
fluence on the detection systems. Together with the flash of particles
accompanying the beam, the capture process produces the bulk of the ionising dose observed 
in the detector system and its electronics.
The tracking  detector \cite{tracker}, composed of $\sim$ 20000 low mass
straw drift tubes, measures the momenta of the charged particles by reconstructing their 
trajectories with the detected hits. Full simulation shows that a momentum resolution 
of O(120 keV) is reached,  thus allowing separation of the CE line from the falling spectrum of the 
DIO  electrons.
The calorimeter plays an important role, complementary to the tracker, by 
providing particle identification capabilities, a fast online trigger filter and
a seed for track reconstruction. In Fig. \ref{caltrack}, the large reduction of
tracking hits correlated with calorimeter timing is shown. The calorimeter should 
also be able to maintain  functionality  in a harsh radiation environment and to work  in the presence of 1 T axial magnetic 
field and in a region evacuated to $10^{-4}$ Torr.

\begin{figure}[!t]
\centering
\includegraphics[width=3.2in]{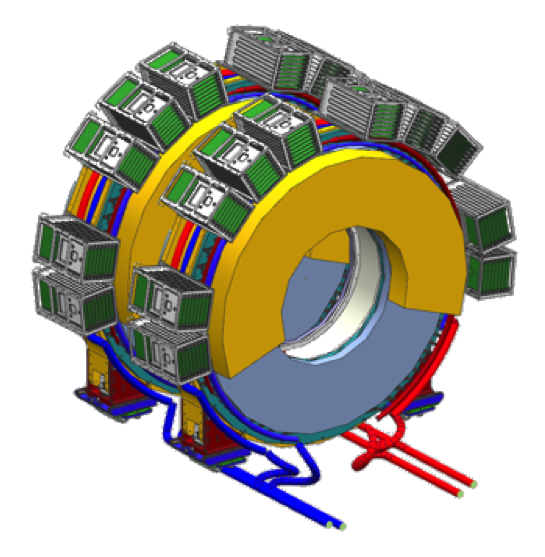}
\caption{CAD of the two calorimeter disks. The crystals are piled up inside the aluminum rings, the light blue area
represents the support for the SiPMs and FEE boards, the yellow area is the envelope of the cables from the FEE to
the digitization system that is located in custom boards inside the crates surrounding the disks.}
\label{calo_cad}
\end{figure}

\begin{figure}[!t]
\centering
\includegraphics[width=2.5in]{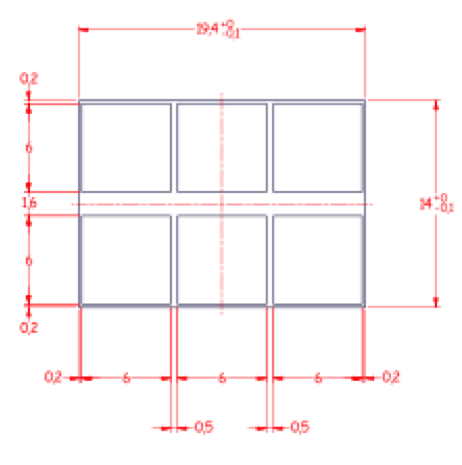}
\caption{Schematic design of the Mu2e custom SiPM array. Each array is composed by 
the parallel of two series of three  6$\times$ 6 mm$^2$ SiPMs. }
\label{mu2e_custom_sipm}
\end{figure}

\section{Calorimeter requirements, technical specifications and calorimeter design}
The tasks to be fulfilled by the calorimeter system translate to achieving the following
requirements for 105 MeV electrons:
(a) a large acceptance;
(b) a time resolution better than 0.5 ns;
(d)  an energy resolution $< 10\%$  and
(d)  a position resolution of 1 cm.

\subsection{Technical choices} 
In order to satisfy these requirements, we have opted for a high quality crystal
calorimeter with Silicon Photomultipliers (SiPMs) readout and with a
geometry organised in two annular disks (Fig. \ref{calo_cad}) to maximise acceptance
for spiralling electrons. The crystals should provide a high  light yield of at least 40 photo-electrons 
(p.e.)/MeV. To reconstruct the events with a high pile-up of particles, fast signals are required
for the combined choice of crystals and SiPMs, thus reducing the selection of
crystals to ones with a decay time ($\tau$) $ < $~40 ns. The Front End Electronics (FEE)  should
provide fast amplified signals to be sampled at 200 Msps (5 ns binning) by the digitisation system.
Selected crystals  (SiPMs) should also be able to sustain a dose of 450 Gy (200 Gy) and a neutron fluence 
of  of $3$ $ (1.2)  \times 10^{12}$ n$/$cm$^2$ while satisfying the calorimeter performance.
Moreover, in order to allow operating the detector inside the DS, without interruption for one year, 
a high redundancy is required on the integrated system. This translates onto having two independent
SiPMs, FEE amplifiers and HV regulator chips/crystal as well as having a completely independent
digitisation system for the two readout lines. A simulation demonstrated that the typical Mean Time
To Failure (MTTF) needed to maintain a fully performing calorimeter along the planned three years 
of running is of $\sim$ 10$^6$ hours/component.

\begin{figure}[!t]
\centering
\includegraphics[width=2.4in]{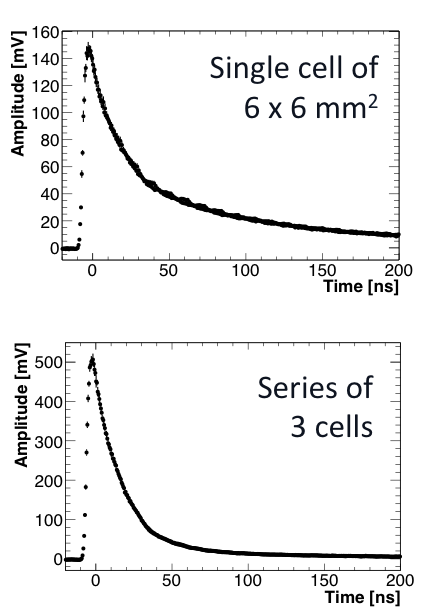}
\caption{ Decay time of the SiPM signal for: (top) a single 6$\times$6 mm$^2$ cell, (bottom) a series configuration of 3 similar cells.}
\label{series_vs_single_sipm}
\end{figure}
After a long R\&D program \cite{calo1,calo2,calo3,calo4}, a final downselect of the scintillating
crystal and photosensor has been done. Undoped CsI crystals were chosen as the best compromise between cost,
performance and reliability. Indeed these crystals are sufficiently radiation hard for our purposes,
have a fast emission time and  a large enough light yield. However, the main scintillation component
is emitted at a wavelength of 310 nm so that,  to well match the  SiPM
Photon Detection Efficiency (PDE) as a function of wavelength, we have selected the new generation 
of UV extended SiPMs. In these sensors, the epoxy resin in the front window has been substituted by a silicon resin
thus providing $> 20 \%$ PDE from the blue region down to 280 nm.
In sect. III.B (Fig. \ref{qa_sipm}),  the PDE measured at 310 nm is shown in good agreement with the vendor specifications.
Since we operate in vacuum, the coupling between
crystals and SiPMs is done without any optical grease (or glue) thus avoiding the related outgassing contribution.
This choice decreased the SiPM light collection efficiency of a factor of two. To recover this  loss, we opted
for a very large area SiPM array. In Fig. \ref{mu2e_custom_sipm}, the design of the Mu2e custom SiPM configuration is shown,
consisting of a 2$\times$3 array of 6$\times$6 mm$^2$ SiPMs, i.e. a parallel of 2 series of 3 sensors. This configuration choice
has been done to reduce the overall array capacitance and the quenching time of the array while simplifying 
the FEE design. This is shown in Fig. \ref{series_vs_single_sipm}, where the quenching time for the series of 3 cells is 
compared with the one of a single cell. As expected the SiPM signal of the series configuration is much narrower than that  of a single cell.
\begin{figure}[!t]
\centering
\includegraphics[width=3.6in]{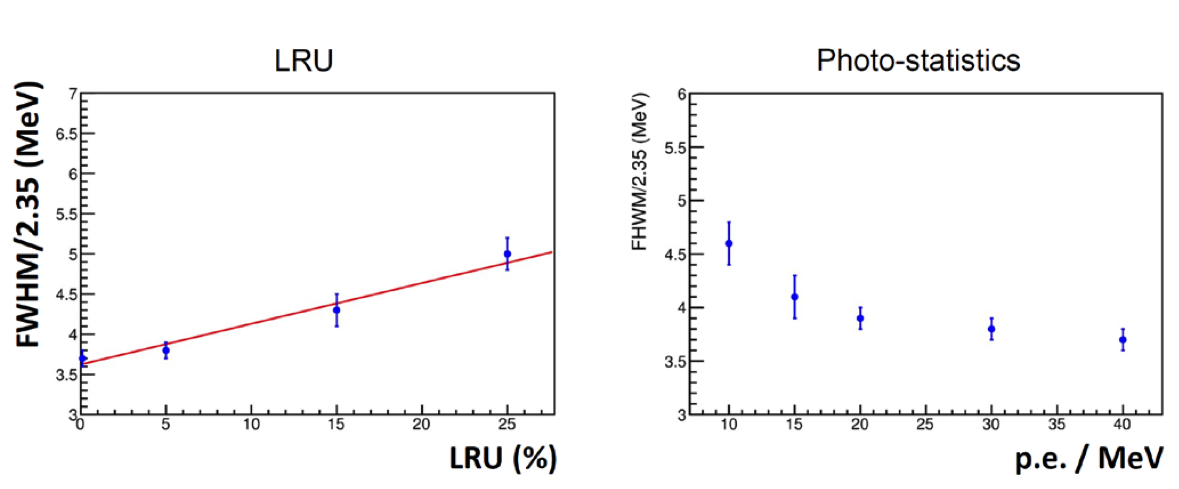}
\caption{Full simulation of the Mu2e calorimeter performance. Dependence of the energy
resolution on: (left) the longitudinal response
uniformity along the crystal axis and (right) the number of photoelectrons/MeV.}
\label{simulation}
\end{figure}

\subsection{Calorimeter design}
A CAD of the calorimeter system is  shown in Fig. \ref{calo_cad}.  
The calorimeter consists of two annular disks, each one made of 674 un-doped CsI
crystals of  parallelepiped shape and dimension of 3.4$\times$ 3.4 $\times$ 20 cm$^3$. The 
calorimeter crystal stack has an inner (outer) radius of 37.4 (66.0) cm; the central hole allowing
the residual muon beam, the beam flash and most of the DIO electrons to pass through it without interacting.
The granularity has been
selected for optimising readout and the digital data throughput. The length of the crystals is
only 10 X$_0$ but is sufficient to contain the 105 MeV electron showers since the electrons
impinge on the calorimeter surface with a $\sim$ 50$^{\circ}$ impact angle. Each crystal is
read by two Mu2e custom SiPMs. Each photosensor
is amplified and regulated in bias voltage by means of
a FEE custom chip. The digitisation system at 200 Msps is located in the nearby 
crates.

The crystal by crystal equalization is obtained by means of a calibration system,
formerly devised for the BaBar calorimeter \cite{babar_calibration},  where a 6.13 MeV photon line is obtained
from a short-lived  $^{16}$O transition. The decay chain comes from the Fluorinert$^{TM}$ coolant
liquid that is activated by  fast neutrons. The activated liquid circulates in aluminium tubes
positioned in the front calorimeter plate to uniformly illuminate each crystal face. The source
system is accompanied by a laser monitoring system that provides a continuous monitoring
capability of the sensor gains and timing offsets, while offering a simple method
to monitor variations also on the energy and timing resolutions.
Usage of cosmic ray and DIO events is foreseen for a continuous in-situ calibration during running.

A full report of the calorimeter design can be found in \cite{FDR}. In the following we extract
the most relevant information for a consistent explanation while reporting the preliminary
measurements done with Module-0.
\subsection{Test of calorimeter performance}
\label{calo_performance}

Before completing the engineering design and starting production, we have tested 
the quality of our design by performing a full simulation of the calorimeter response 
and a test beam on a small size prototype.

A  calorimeter simulation of CE events has been performed with Geant4, in the Mu2e software framework. This simulation
includes: (i) the overlap inside the event (pile-up) of the accidental background activity coming from DIO 
and muon capture events,  (ii) a full description of the signal shapes related to the convolution of scintillation, SiPM and FEE response and (iii) 
a description of the main elements of the mechanical structure. As an example, in Fig. \ref{simulation}, the
energy resolution dependence on  light yield (LY), expressed as  the number of photoelectrons (p.e.)/MeV, and on the longitudinal
response uniformity (LRU) are shown. The Mu2e calorimeter design looks robust against variation of the 
parameters providing  resolution  better than 5\%. However, the optimal working point is
observed for LY $>$ 20 p.e./MeV for each SiPM and a LRU $<$ 5\%.

A calorimeter prototype consisting of a $3 \times 3$ matrix of $30 \times 30 \times 200$ mm$^2$  un-doped CsI crystals 
wrapped with 150 $\mu$m Tyvek, each read out by one $12 \times 12$ mm$^2$ SPL TSV Hamamatsu MPPC,
has been tested with an electron beam at the Frascati Beam Test Facility (BTF) in
April 2015.  The final choice of calorimeter wrapping was done after a test of alternatives (Teflon, Tyvek and Aluminum
foils) by considering their light yield collection, their radiation hardness  and their easiness  on the wrapping technique.
The best light collection was obtained by means of 8 layers of 20 $\mu$m Teflon, its difference with 150 $\mu$m 
Tyvek (Al) foils was at the level of +10 (25) \%. All options were good for radiation resistance. We finally opted
for Tyvek for reproducibility in the wrapping method.

The test beam results, described in \cite{calo3}, are consistent with a dedicated Geant-4 simulation and provide:
\begin{itemize}
\item a time resolution better than 120 ps for 100 MeV electrons,  ranging from about 250 ps
at 22 MeV to about 120 ps in the energy range above 50 MeV;
\item an energy resolution of $\sim7\% $ for 100 MeV electrons, dominated by shower leakage and beam energy spread.
\end{itemize}

\begin{figure}[!t]
\centering
\includegraphics[width=3.6in]{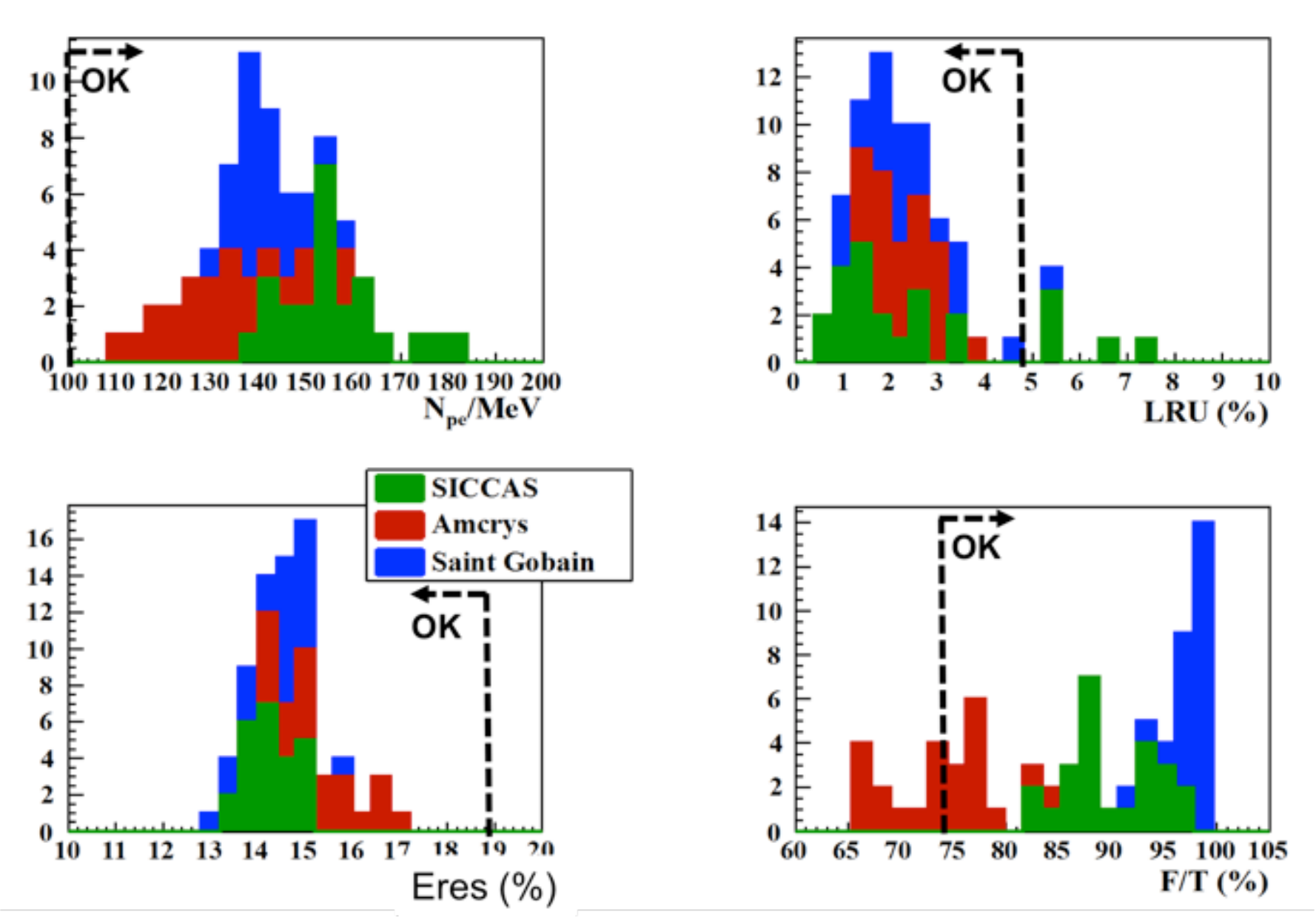}
\caption{Quality assurance of optical properties for the pre-production crystals.   }
\label{preprod_crystals}
\end{figure}

\begin{figure}[!t]
\centering
\includegraphics[width=3.0in]{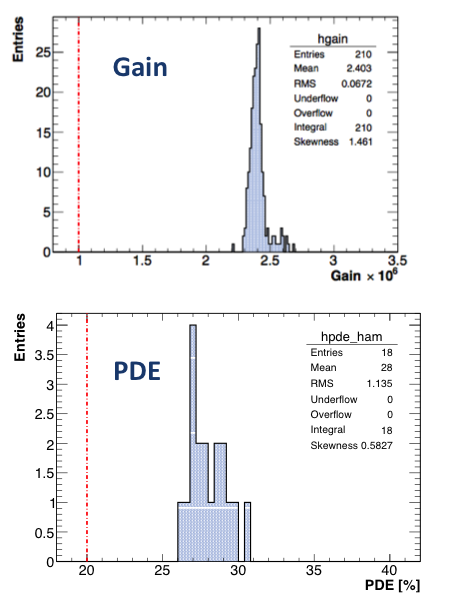}
\caption{Example of QA for the pre-production Hamamatsu Mu2e SiPMs: (top) distribution
of the SiPM gains, (bottom) distribution of the PDE. Red dashed lines indicates
the minimum acceptable value.}
\label{qa_sipm}
\end{figure}

\begin{figure*}[!t]
\centering
\subfloat[Case I]{\includegraphics[width=3.0in]{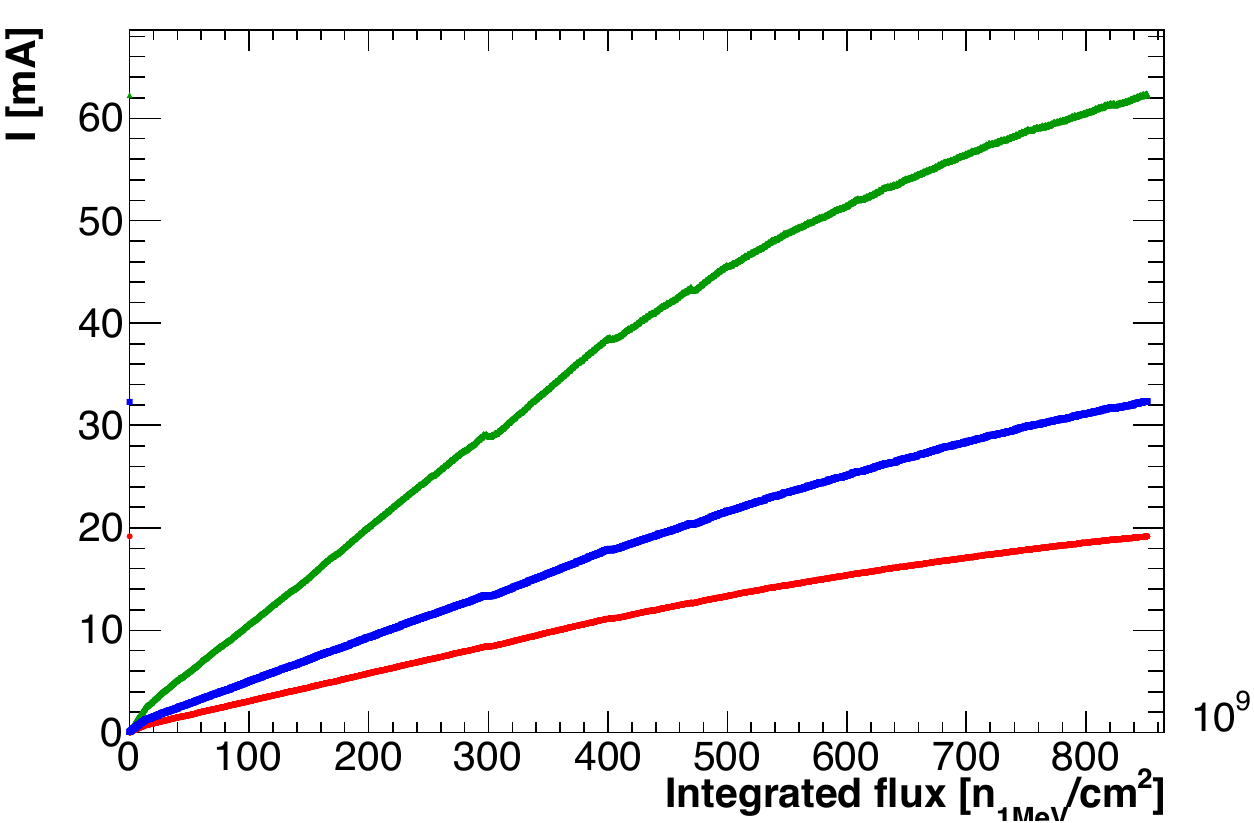}
\label{fig_first_case}}
\hfil
\subfloat[Case II]{\includegraphics[width=3.0in]{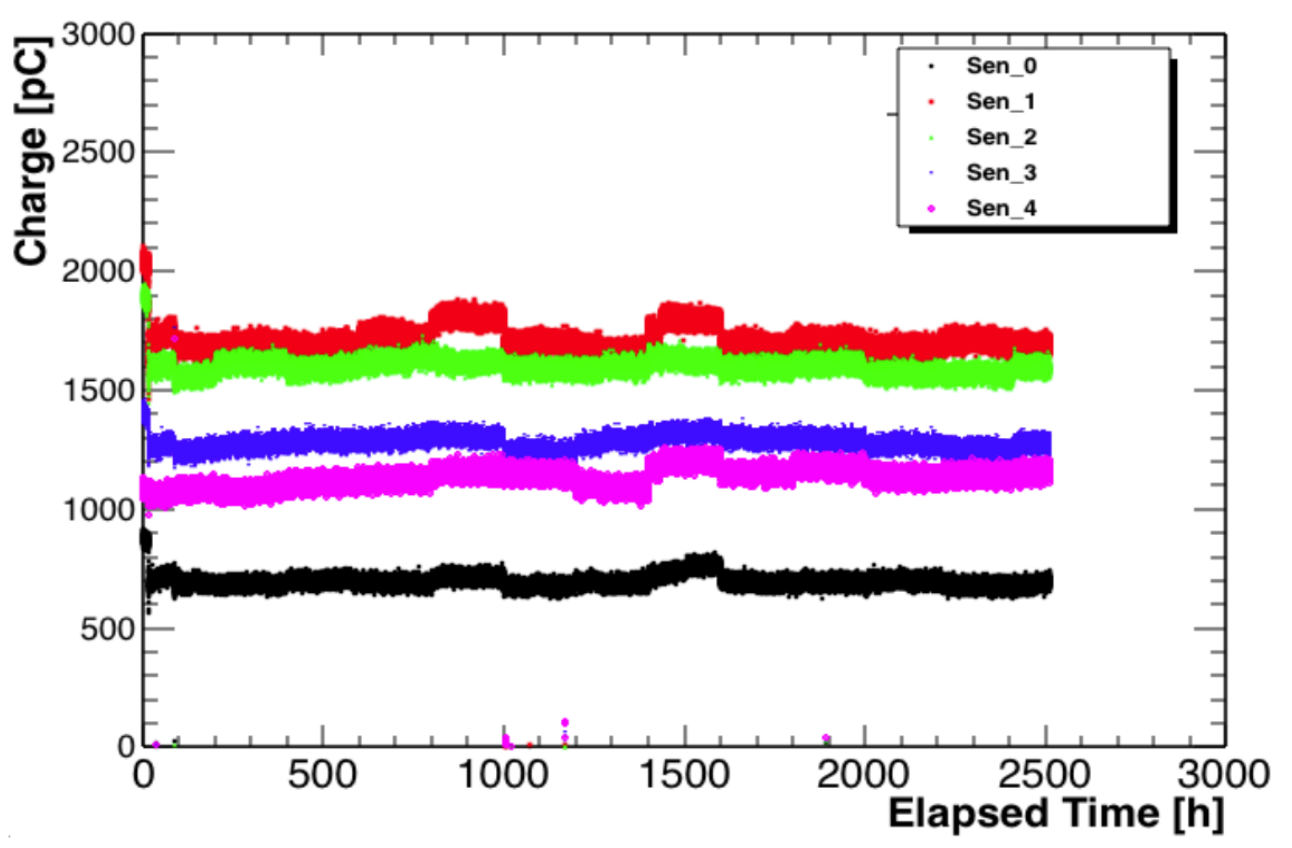}
\label{fig_second_case}}
\caption{Quality control of the pre-production SiPMs: (Left) Dependence of the leakage current on the total
neutron fluence,  for neutron of 1 MeV equivalent energy. Green, blue and red curves are for SensL, Advansid and Hamamatsu SiPMs respectively. (Right) charge response to a LED illumination as a function of running time for the
SiPMs undergoing an MTTF test while kept in an oven at 50 $^{\circ}$C.}
\label{mttf_and_irradiation}
\end{figure*}

\section{Status of preproduction}
After freezing the design, in 2016 we have carried out a pre-production of the basic components to control specifications
and select vendors for the final production phase. In the following sections, we describe pre-production of crystals,
SiPMs and electronics.

\subsection{Pre-production of crystals}
A total of 72 crystals, of final dimension and shape, have been procured from three vendors: St. Gobain (France), SICCAS (China) and
Amcrys (Ukraine). All crystals have been exposed to a very accurate Quality Assurance phase. As soon
as received at Fermilab, a test of their mechanical precision, both on dimension and shape, has been carried out at the Coordinate
Measuring Machine of  the SiDET facility. We have then measured (both at Frascati and Caltech) their optical quality by determining,
for each crystal, the light yield, the LRU, the energy resolution at 511 keV and the Fast/Total component (i.e., the ratio of the energy
seen integrating the signal in 200 ns with respect to the total signal integrated in 3 $\mu$s). Fig. \ref{preprod_crystals} shows
a summary of the crystals' optical properties. The LY measurement have been carried out with UV extended  photomultipliers
in order to improve the resolution of the 511 keV peak and to simplify the comparison with the quality control from the producers.
A test of the radiation induced noise and of the radiation hardness to
dose and neutron fluence has also been carried out. A full report on these measurements can be found in
another contribution to these proceedings \cite{zhu}. At the end of the pre-production phase, we have decided
to use both St. Gobain and SICCAS firms for the production crystals, in order to minimise schedule risks.
 
\begin{figure}[!t]
\centering
\includegraphics[width=3.6in]{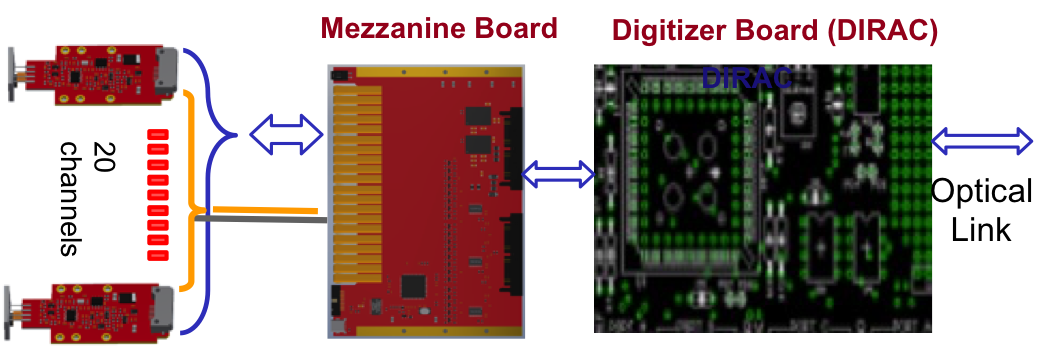}
\caption{Electronics scheme: groups of 20 SiPMs  are connected to the Mezzanine board and digitized by the DIRAC board that sends data to TDAQ.}
\label{FEE_scheme}
\end{figure}

\begin{figure}[!t]
\centering
\includegraphics[width=2.0in]{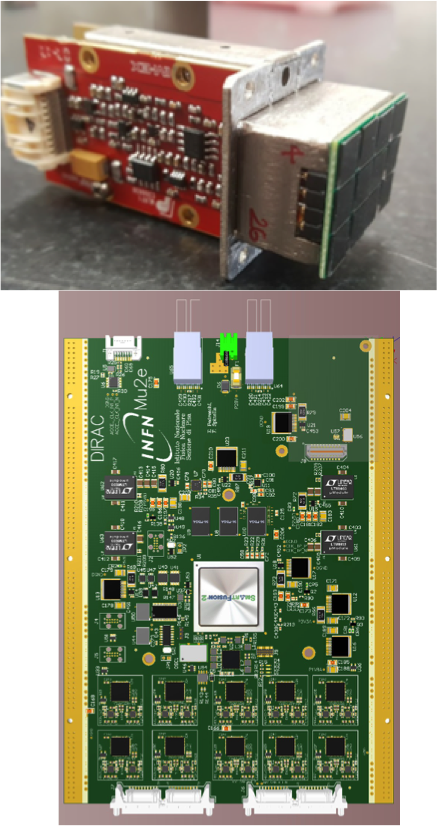}
\caption{Picture of the pre-production electronics: (top) example of two Amp-HV chips
connected to the back of the Left and Right SiPM arrays on their holder; (bottom) the first version
of the calorimeter digitizer (DIRAC).}
\label{electronics}
\end{figure}

\subsection{Pre-production of SiPMs}
A total of 150 Mu2e SiPMs   has been procured from three vendors: Hamamatsu (Japan), SensL (Ireland) and Advansid (Italy).
All received SiPMs have been exposed to a very accurate Quality Assurance phase.  As soon
as received in Pisa, they have been tested in a semi-automated station for their basic properties.
These SiPMs have been kept to a constant temperature of 20 $^{\circ}$C by means of a Peltier cell and a copper
line. All measurements were performed for each of the six cells in the array. For each cell, we measured the  I-V curve, determined
the breakdown voltage, V$_{\rm br}$  and derived the operating voltage, V$_{\rm op}$, as 3 volts above V$_{\rm br}$.
For each SiPM array,  the I$_{\rm dark}$ spread at the operating voltage 
has also been determined. The SiPM gain of each cell was evaluated by means of a photon-counting technique, at the 
chosen V$_{op}$, with a high gain amplifier.   Fig. \ref{qa_sipm}.top shows the gain distribution for all pre-production 
Hamamatsu SiPMs. Finally, the photon detection efficiency (PDE) of each cell was determined by illuminating the sensors 
with a pulsed LED and extracting the number of events with N$_{\rm pe}=0$ and with N$_{\rm pe} \geq $ 1. The result has been 
then cross-calibrated with the known PDE of a reference SiPM. The PDE distribution obtained
for  the Hamamatsu SiPMs is shown in Fig. \ref{qa_sipm}.bottom.

The SiPMs were also tested for their radiation hardness by exposing them both to an ionizing dose and
to a neutron fluence. The total irradiation dose  (TID) was delivered at the Calliope facility of ENEA Casaccia, Italy.
A single sensor was exposed to a high intensity $^{60}$Co source integrating  a TID of 200 Gy, that is
our requirement limit.
A small increase in the leakage current has been observed with  negligible effect on the
SiPM response. The test with neutrons was carried out at the EPOS facility in HZDR, Dresden, Germany.
In this facility, a 30 MeV electron beam, of O(100 $\mu$A) current,  interacts with 1 cm thick tungsten
target and becomes a good source of photons and neutrons. The target is shielded with  lead and
borated polyethylene. Above the shielding roof, a clean neutron flux centered
at 1 MeV, with negligible photon contamination, is available for testing. One SiPM per vendor has been
exposed to a fluence up to 8.5 $\times 10^{11}$ n(1 MeV)/cm$^2$.
A linear increase of the dark current as a function of the fluence is observed, with a different
slope between vendors (see Fig. \ref{mttf_and_irradiation}.left).
In the hottest calorimeter places, we expect to integrate, during the experiment, a fluence
of up to 1.2$\times$10$^{12}$ n(1 MeV)/cm$^2$ (5 years of running and a simulation
safety factor of 3). In the case of the best vendor and after annealing, this irradiation level 
corresponds to  a dark-current of 30 mA, i.e. a 
power consumption of 4.5 W/SiPM  that makes the SiPMs unusable. The solution is to 
limit the  current drawn up to a maximum acceptable value of 2 mA by using two
handles: (i) cool down the SiPM at a running temperature of $\sim$ 0 $^{\circ}$C and 
(ii) apply a reduction in operating voltages. 

A  measurement of the  MTTF for the different SiPMs has also been carried out.
A group of 5 pieces/vendor has been stressed for 3.5 months by operating them at V$_{op}$ 
inside a light tight box kept at a temperature of 50 $^{\circ}$C. During this stress-test period,
 the sensors were continuously monitored by: (i) controlling their response to a pulsed
 LED fired every 2 minutes and (ii) registering the behaviour of the dark current  with time. All
 SiPMs under test were still alive and perfectly working at the end of the stress period. Assuming an
 acceleration factor of 100 (due to the difference between experiment and stress temperatures) 
 we have estimated an MTTF larger than 6$\times 10^5$ hours for each vendor. This first determination of MTTF 
 already grants a safe running condition for the first two years of running. A more dedicated measurement 
 will be carried out during the SiPMs production phase.
\subsection{Preproduction of electronics}
 The front-end electronics (FEE) consists of two discrete and independent chips (Amp-HV), for each crystal, directly connected to the 
 back of the SiPM pins. These provide the amplification and shaping stage, a local linear regulation of the bias voltage, monitoring of current and temperature on the sensors and a test pulse. In Fig. \ref{electronics}.top, an example of the left/right FEE chips inserted in the SiPM holder
 is shown. For equipping the Module-0 (see sub-sec \ref{sec:module0}), a first pre-production of 150 FEE chips has been carried out. A second version will be produced to tune the amplification value and the shaping section, after completing the analysis of the test beam data.
 
 Each disk is subdivided into 34 similar azimuthal sectors of 20 crystals. 
Groups of 20 Amp-HV chips are controlled by a dedicated mezzanine board (MB), where an ARM controller distributes the LV and the HV reference values, while setting and reading back the locally regulated voltages.  Groups of 20 signals are sent differentially to a digitizer module (DIRAC,
 DIgitizer and ReAdout Controller) 
 where they are sampled (at 200 Msps) and processed before being optically transferred to the T-DAQ system. The Detector Control System (DCS) parameters, read out/set by the MB, are passed via I2C to the DIRAC boards that then communicate them to the Mu2e DCS system through an optical link. 
 In Fig. \ref{electronics}.bottom, a picture of the version-0 of the  DIRAC board  is shown.

\section{Engineering design}
\begin{figure}[!t]
\centering
\includegraphics[width=3.2in]{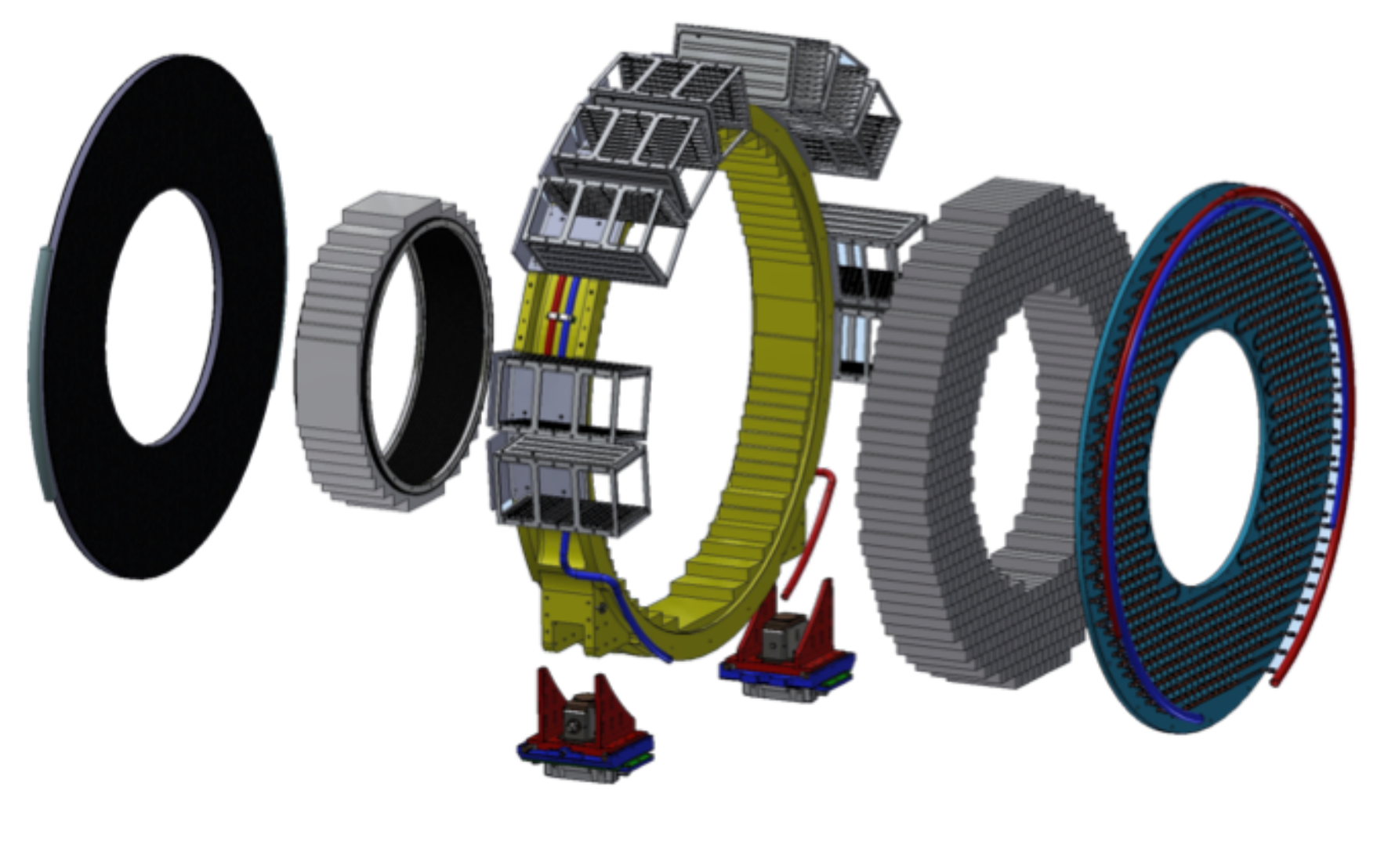}
\caption{Exploded view of calorimeter mechanics: from left to right we distinguish the carbon fiber front plate,
the inner ring, the aluminum disk support with crates and feet, the crystals, the front-end disk with cooling
lines where SiPMs and FEE will be inserted.}
\label{calo_mechanics}
\end{figure}

Figure~\ref{calo_mechanics} shows an exploded view of a single calorimeter annulus.
It consists of an outer monolithic Al cylinder that provides the main support for the crystals and integrates the feet and adjustment mechanism to park the detector on the rails inside the detector solenoid. The inner support is made of a carbon fiber  cylinder that maximises $X_0$, i.e. minimises the passive material.
The crystals are then sandwiched between two cover plates.  A front plate in  carbon fiber intercepts the electrons coming from target. It also integrates 
the thin wall Al pipes of the source calibration system to flow the Fluorinert$^{TM}$.  A back  plate, made of PEEK, with apertures in correspondence 
of each crystal, is used to lodge  the FEE and SiPM holders. The back plate houses also the copper pipes where a coolant is flowed to thermalise the photosensors to low temperature and extract the power dissipated by both FEE and sensors. Ten custom-made crates are arranged on top of the outer cylinder and are connected to the cooling circuit to cool the digitizer boards.

A full scale mock-up of the mechanical structure is being built to test the assembly of the crystals, FEE electronics, cooling system and the
overall structure robustness. The Al outer ring, the inner Carbon Fiber cylinder, quadrant sections of the front and back plates and one crate 
have already been manufactured. A whole annulus will be assembled using a mixture of fake iron crystals and a sample of pre-production CsI crystals. 

\section{Design, assembly and test of Module-0 }
\label{sec:module0}
\begin{figure}[!t]
\centering
\includegraphics[width=3.2in]{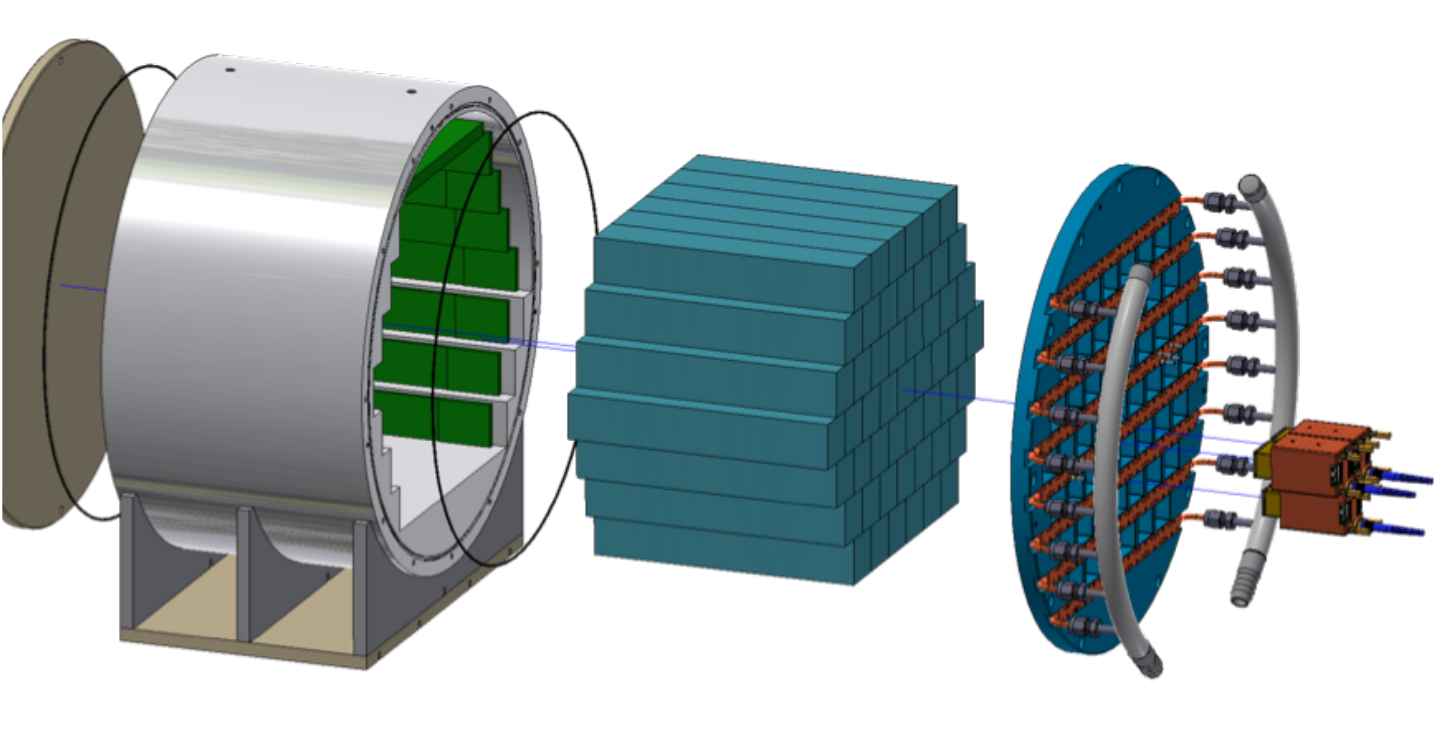}
\caption{Breakdown of Module-0 components. From left to right, we distinguish the front plate, the aluminum disk, the crystals,
 the front-end disk with cooling lines and two examples of the SiPM and FEE holders.}
\label{module0}
\end{figure}

A large size prototype, dubbed Module-0, was built using  components of final size and dimensions to resemble a portion
of a final disk and to test all final elements for crystals, sensors, mechanics and electronics. Fig. \ref{module0}  
shows an exploded view of the Module-0 CAD drawings to present the various components. This detector is composed by
51 CsI pre-production crystals, each one instrumented with two pre-production Mu2e custom 
SiPMs.  The first version of SiPM copper holders and FEE chips (Fig. \ref{electronics}.top) was  used. 
This allowed to test the gluing procedure of SiPMs on the copper holders, the pre-amplification and HV regulation schemes  and the cooling performance.
The cooling and FEE support disk was built, as in the final detector, with an insulated support milled to lodge
both the copper cooling lines and the SiPM/FEE holders. The only difference with respect to the final version is the insulating
material used  (Zedex instead of Peek). 

The prototype was equipped with  two FEE chips/crystal in the central
crystal and in the surrounding ring, and with one FEE chip/crystal in
all other rings. The readout was organised in five mezzanine
boards (16 channels/each), in NIM format, to set and read HV and temperature
for all SiPMs. Digitisation was  performed with two commercial boards from CAEN,
each with 32 channels of 1 GHz digitisers (DRS4).

\subsection{ Test beam results}
\begin{figure}[!t]
\centering
\includegraphics[width=2.8in]{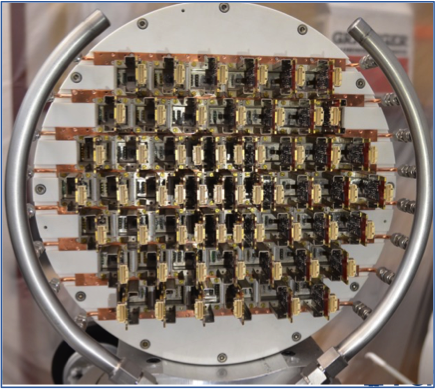}
\caption{Picture of Module-0: FEE/SiPM holders are clearly visible.}
\label{module0_picture}
\end{figure}

\begin{figure}[!t]
\centering
\includegraphics[width=2.8in]{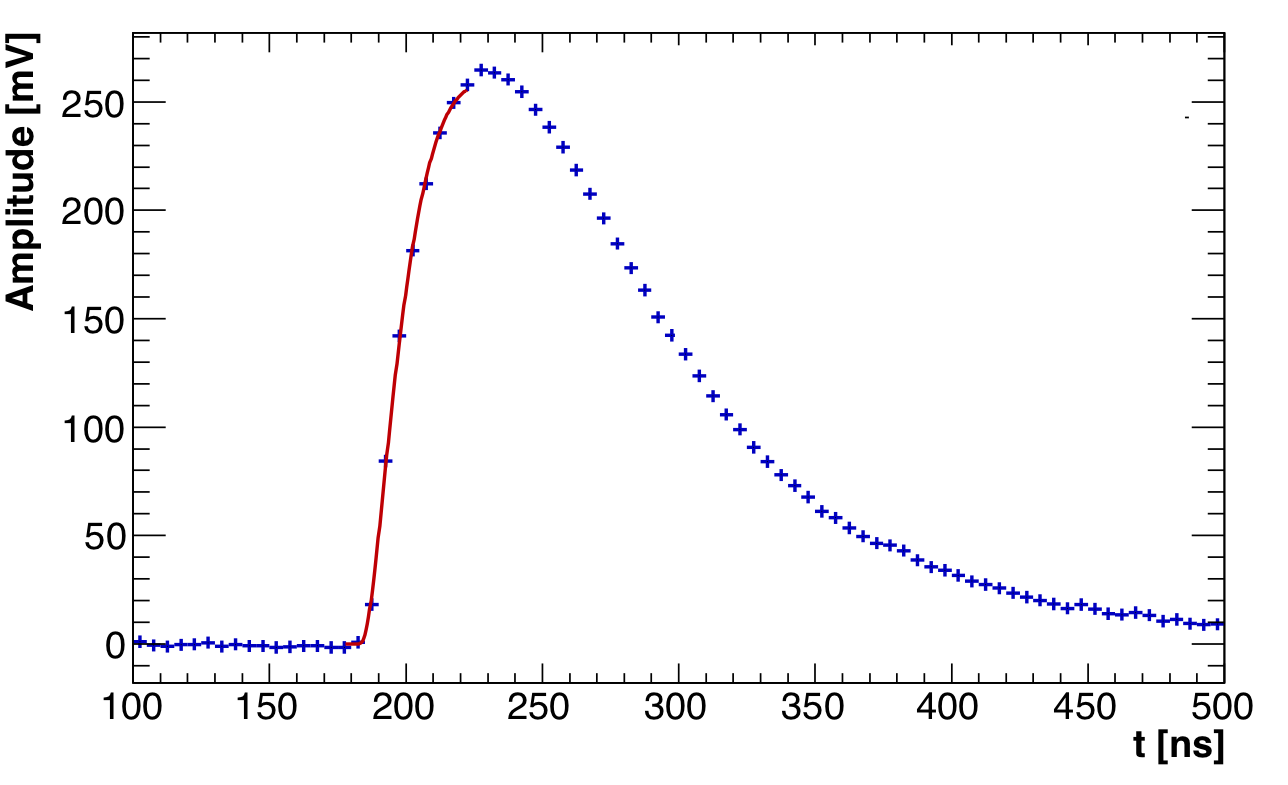}
\caption{Pulse shape of the central crystal for 100 MeV electrons.}
\label{pulse_height}
\end{figure}

\begin{figure*}[!t]
\centering 
\subfloat[Energy reconstructed in the whole calorimeter.]{\includegraphics[width=2.6in]{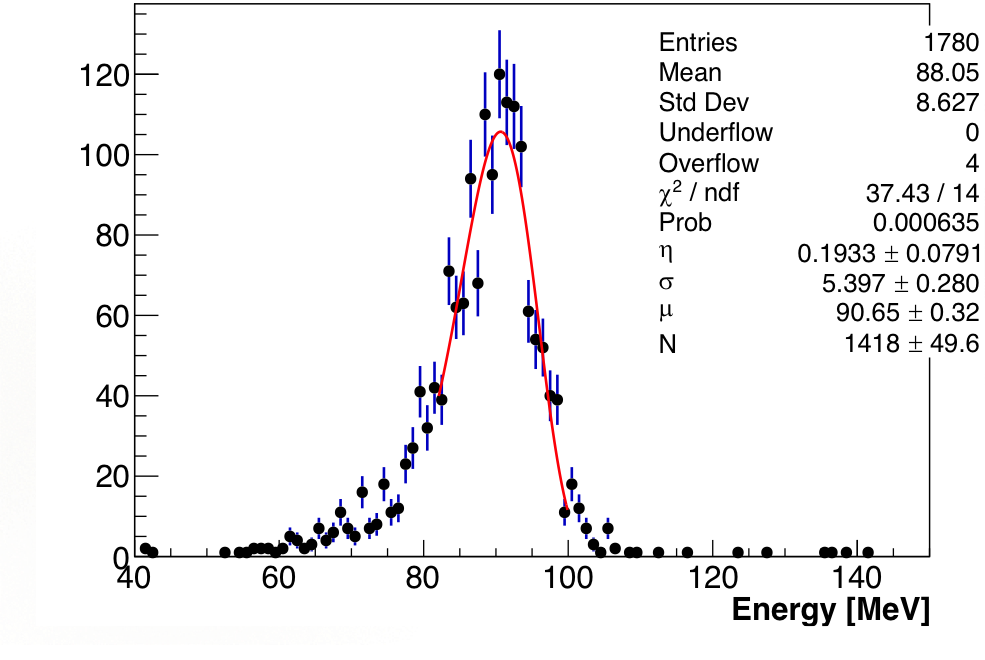}
\label{fig_first_case}}
\hfil
\subfloat[Timing difference between the two central crystal SiPMs.]{\includegraphics[width=2.8in]{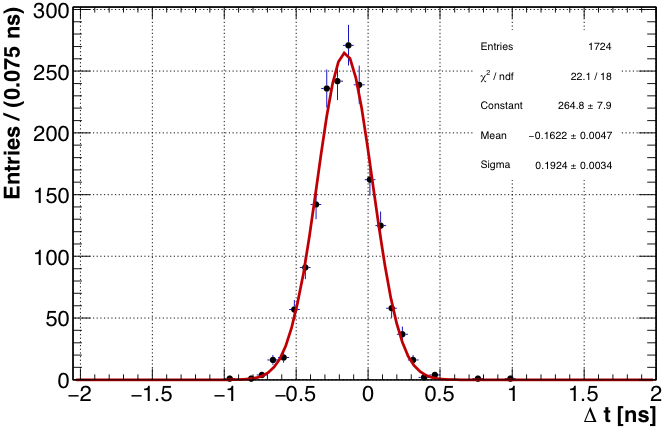}
\label{fig_second_case}}
\caption{Preliminary results  of 100 MeV electron beams for the Module$\_$0 test beam.}
\label{module0_testbeam}
\end{figure*}

The Module-0 was exposed to an electron beam with energy between 60 and 120 MeV at the
Beam Test Facility (BTF) of Laboratori Nazionali di Frascati of INFN, in the week 8-15 of May
2017.  The system integration was really successful. The cooling system was 
properly working allowing us to keep the sensors' temperature 
at (20 $\pm$ 1)$^\circ$C.  
We collected few scans in energy and position with beam at normal incidence 
and some special runs at 50 degrees to simulate the impact angle we will experience in
the Mu2e experiment.  

In Fig. \ref{pulse_height}, the pulse height observed in the central crystal for
100 MeV electrons is shown. The rise-time of the signal is of around 30 ns, while
the convoluted decay time of scintillation light, SiPM response and FEE shaping
has a $\tau$ of $\sim$ 50 ns, for a total signal width of $\sim$ 350 ns. The charge of
each channel is obtained, after subtracting the baseline, with a signal integration time
of 250 ns.

In Fig. \ref{module0_testbeam}.a, the energy distribution for 100 MeV electrons at normal incidence is shown. 
This was obtained summing up the charge of all calorimeter channels with an energy deposition larger than 5 $\times \sigma_{\rm noise}$,  
that represents the noise level determined with a Gaussian fit to the response observed without beam. 
Channel by channel equalization was done reconstructing the energy deposited by cosmic ray events selected
 with external scintillation counters. This calibration is in good agreement with an independent determination 
 based on samples of 100 MeV electrons sent in the center of each crystal in the two innermost rings. 
The energy plot has been fit with a log-normal function as shown 
 by the red curve overimposed. An energy resolution better than 6\% is obtained, thus improving the result found 
 with the small size prototype and satisfying the calorimeter requirement.


The timing resolution of the calorimeter was also
investigated. The rise time of each signal shape was fit to determine
the crossing time, as shown in Fig.\ref{pulse_height}. The distribution of
the timing difference between two SiPMs of the central crystal is shown
in Fig. \ref{module0_testbeam}.b for 100 MeV electrons. 
A timing resolution of 96 ps is estimated when considering the semi-difference of the two times.
This estimate takes into account the resolution related to photoelectron statistics and
readout electronics but it does not consider the jitter due to shower fluctuation or
clock synchronisation. Further measurements are planned to clarify these contributions.

\section{Conclusions and perspectives}
The Mu2e calorimeter is a state of the art crystal calorimeter with excellent
energy ($ < 10$\%) and timing ($<$ 500 ps) resolutions, for 100 MeV electrons,
and a good pileup  discrimination capability. The latter one is obtained  thanks to the chosen crystals 
and sensors assisted  by a fast analog electronics and a digitisation at 200 Msps. 
There are many demanding 
requests to be satisfied by this detector, such as to keep the required performance  
in presence of 1  T axial magnetic field, in an evacuated  region and in a radiation harsh 
environment. The CsI crystals will withstand  the expected dose and fluence  with a small 
light yield loss. The Mu2e SiPMs will work under neutron irradiation when cooled to 
0 $^{\circ}$C, thus asking for  a good engineering design of the calorimeter mechanics 
and of its cooling system.
Pre-production of crystals and SiPMs has been successfully carried out
in 2016 demonstrating that the required technical specifications can be met.
Final vendors have been selected and the production will start at the end of 2017.
Pre-production for the electronics is also under way. The calorimeter Module-0 has 
been built in April 2017 to exercise the final selected components and the engineering
of mechanical and electronic systems. A full size mock-up of the calorimeter disk
is underway to test the assembly procedure. The schedule is to start assembly the first 
disk in winter 2018 and complete the calorimeter construction in 2020.

\section*{Acknowledgment}
We are grateful for the vital contributions of the Fermilab staff and the technical staff of the participating institutions.
This work was supported by the US Department of Energy; 
the Italian Istituto Nazionale di Fisica Nucleare;
the Science and Technology Facilities Council, UK;
the Ministry of Education and Science of the Russian Federation;
the US National Science Foundation; 
the Thousand Talents Plan of China;
the Helmholtz Association of Germany;
and the EU Horizon 2020 Research and Innovation Program under the Marie Sklodowska-Curie Grant Agreement No.690385. 
Fermilab is operated by Fermi Research Alliance, LLC under Contract No.\ De-AC02-07CH11359 with the US Department of Energy, Office of Science, Office of High Energy Physics.
The United States Government retains and the publisher, by accepting the article for publication, acknowledges that the United States Government retains a non-exclusive, paid-up, irrevocable, world-wide license to publish or reproduce the published form of this manuscript, or allow others to do so, for United States Government purposes.



%

\end{document}